# Methods to Measure the Broncho-Arterial Ratio and Wall Thickness in the Right Lower Lobe for Defining Radiographic Reversibility of Bronchiectasis

**Abhijith R. Beeravolu[1], Ian Brent Masters[3,4], Mirjam Jonkman[1], Kheng Cher Yeo[1], Spyridon Prountzos[5], Rahul J Thomas[3,4], Eva Ignatious[1], Sami Azam[1], Gabrielle B McCallum[2], Efthymia Alexopoulou[5], Anne B Chang[2,3,4], Friso De Boer[1]**

[1]Faculty of Science and Technology, Charles Darwin University, Casuarina, Northern Territory, Australia
[2]Child and Maternal Health Division, Menzies School of Health Research, Darwin, Charles Darwin University, Darwin, Northern Territory, Australia
[3]Department of Respiratory and Sleep Medicine, Queensland Children's Hospital, Brisbane, Queensland, Australia
[4]Australia Centre for Health Services Innovation, Queensland University of Technology, Brisbane, Queensland, Australia
[5]Second Department of Radiology, Faculty of Medicine, Attikon University General Hospital, National and Kapodistrian University of Athens, Athens, Greece

Corresponding author: Eva Ignatious (e-mail: eva.ignatious@cdu.edu.au).

**ABSTRACT** Bronchiectasis is a chronic respiratory disorder characterized by dilated and damaged bronchial walls resulting from recurrent and protracted episodes of inflammation and infection. The diagnosis of bronchiectasis requires objective measurement of abnormal bronchial dilation. It is confirmed radiologically using a chest C.T. scan where the pathognomonic feature is increased broncho-arterial ratio (BAR) (>0.8 in children) and, often accompanied by other features such as bronchial wall thickening. Developing image processing-based methods facilitates quicker interpretation of the studies and detailed evaluations according to the lobes and segments. However, there are many challenges, such as inclined nature, oblique orientation, and partial volume effect, which can make it challenging to obtain accurate measurements of structures in the upper and middle lobe using the same algorithms. Thus, the accurate detection and measurements of airway and artery regions for BAR and wall thickness for each lobe require the development of different image processing/machine learning methods and approaches. Here, we adopt a step-by-step approach and propose methods for three steps: 1. Separating right lower lobe (RLL) region from full-length C.T. scans using tracheal bifurcation (Carina) point as a central marker; 2. Updated technique to locate inner diameter of airways and outer diameter of arteries for BAR measurement; and 3. Measuring airway wall thickness (W.T.) by identifying the outer and inner diameter of airway boundaries (perimeter). Our analysis of 13 HRCT scans with varying thicknesses (0.67mm, 1mm, 2mm) demonstrates that the frame containing the tracheal bifurcation can be detected accurately in most cases, with a deviation of ±2 frames in some cases. Similarly, a Windows app is developed for measuring inner airway diameter, artery diameter, BAR, and wall thickness, allowing users to draw bounding around visible discrete B.A. pairs in the RLL region. Measurements of 10 B.A. pairs revealed accurate results comparable to those of a human reader, with deviations of ±0.10-0.15mm observed across all measurements. Additional studies and validation are necessary to consolidate inter- and intra-rater variability and enhance the proposed methods.

## I. INTRODUCTION

The global population stands at 8 billion this year and is anticipated to reach 11 billion by the end of this century [1]. Over time, healthcare has evolved from primitive and spiritual practices to more systematic and scientific approaches. Each era is built upon previous knowledge, highlighting the importance of continuous innovation and knowledge exchange in improving healthcare outcomes [2][3]. The growth of noncommunicable diseases (NCDs) worldwide has become a significant public health concern driven by demographic, behavioral, environmental, and socioeconomic factors. Among these, bronchiectasis stands out as a chronic lung condition that is increasingly recognized yet often neglected across low, middle, and high-income countries [4-7], with considerable inequality between and among countries [8]. Although the number of healthcare professionals is growing worldwide, this growth might not keep pace with the projected increase in the world population [9], as most of this growth is only seen in high-income countries to meet the demands of increasing chronic diseases [10][11]. It is imperative for developed, developing, and underdeveloped nations to collaborate on global initiatives that unify healthcare professionals under common standards [12][13]. This would enable healthcare professionals to move freely between countries, not only in times of crisis but also in regular circumstances [14]. They need to ask the question, are they only looking for development in their economy, or development in their minds? [15][16]. The current study examines bronchiectasis, a chronic condition that requires urgent attention from the global healthcare community, as it often becomes irreversible if neglected and can be reversible if diagnosed and treated early in childhood [17][18]. Cohort Australian and English studies [19][20] on adults with symptoms from childhood have reported that the disease is worse in adults who were symptomatic from childhood. Appropriate treatment plans [21][22] and to-and-fro movement of healthcare professionals between nations are essential to prevent the spread of this disease in the airways, which can significantly impact the difference between a healthy, fulfilling life and a burdensome one [23][24]. Although bronchiectasis is conventionally considered irreversible, small studies have demonstrated its reversibility in children, with recent research indicating that younger age and lesser radiographic severity at diagnosis are key factors associated with radiographic reversibility on chest HRCT scans [18][25]. Early diagnosis and treatment are likely to be critical to reverse bronchiectasis, but many countries struggle to achieve this due to economic and health inequalities. These challenges can potentially be mitigated by developing digital technologies by incorporating medical and experiential knowledge from various experts.

The current guidelines recommend using a lower broncho-arterial ratio (BAR) cutoff of >0.8 to define abnormality in children [17][22][26]. The BAR is determined using the ratio between inner airway and outer artery diameters [22]. For the bronchial-wall thickness, clinicians measure the difference between the outer and inner airway diameters and divide it by two [27][28]. The method operates under the assumption that the bronchial wall is symmetrically distributed around the lumen for simplicity in many clinical and research settings.

Manually inspecting Broncho-Arterial (BA) pairs in HRCT scans can be time-consuming; thus, image processing/machine learning methods could markedly shorten the time process and potentially enhance accuracy. In this research, we proposed methods to detect the inner airway, outer airway, and outer artery boundaries to measure BAR and wall thickness (W.T.). These methods might allow the researchers to measure the BA pairs in the full-length C.T. scans, provide comprehensive airway tree evaluations according to the lobes and segments quickly.

Several research groups have worked on high-resolution computed tomography (HRCT) scans and have proposed methods for extracting the lung structures out to the sub-lobar level and quantitating various anatomical features to evaluate and track the progression of diseases affecting the airways and arteries[29-31]. Multiple studies [32-44] have been conducted using 2D-based methods to detect and measure the B.A. pairs for BAR and wall thickness, and these methods are consistent with human readers. They have used region growing, morphology, full-width-half maximum, phase congruency, pixel-level image analysis, edge-radius-symmetry (ERS) transform, and A.I. model-based methods to segment the airway and artery regions for automated measurements. Researchers have transitioned from 2D to 3D imaging methods to improve the accuracy and reliability of evaluating the structures in the lung parenchyma region. 3D Methods for segmentation, detection, and other tasks have shown promising results in the medical field [45-56], which is the goal of our future research. We believe that straightforward image processing techniques and 3D-based methods can offer substantial insights into diseases and optimize workflows for healthcare professionals more effectively than AI-based systems, which often lack generalizability [57][58]. One could argue that there is a growing proliferation of publications in the medical A.I. field that offer minimal practical benefit to medical professionals [59-61].

Therefore, we aimed to develop methods for analyzing HRCT scans to understand the aspects of bronchiectasis according to lobes and segments using 2D (current) and 3D (future) image processing methods. The goal is to determine the necessary imaging parameters for accurate measurements in each lobe. For example, the BA pairs in the middle and upper lobes are inclined (angled), making it difficult to obtain accurate measurements on axial C.T. slices. Challenges such as oblique orientation and partial volume effect can make the inclined pairs visibly incomplete, blurring the boundaries between the structures in those lobes. Therefore, different image processing and measurement methods are required for each lobe to get accurate measurements. The research starts with the right lower lobe (RLL), focusing on identifying the central marker (i.e., bifurcation point), inner and outer airway, and outer artery regions to extract accurate measurements.

### A. NOVELTY OF THE PROPOSED METHODS IN DIGITAL BRONCHIECTASIS STUDIES

In our previous study [32], we developed methods to locate objects like airways, arteries, and potential bronchiectasis-associated pairs to identify discrete B.A. pairs (DBAs) and measure the BAR in full-length HRCT scans. However, this approach has limitations as it doesn't allow for detailed analysis of bronchiectasis in specific lobes or segments. For better disease evaluation and enhanced confidence in routine clinical use, we should adopt a step-by-step approach to develop methods for understanding each aspect of bronchiectasis rather than analyzing the entire C.T. scan at once. This approach differentiates the current research from our previous and existing studies. The main aim of this study is to separate the frames with information (B.A. pairs) related to the segments RB6, RB7, RB8, RB9, and RB10 from the full-length HRCT scans and measure them for BAR and wall thickness. The novel contributions of our paper are listed below.

- Developed methods to extract the right lower lobe (RLL) region from full-length HRCT scans (0.67mm, 1mm, and 2mm) using Carina (tracheal bifurcation) as a central marker.
- Proposed methods on mediastinum window C.T. scans to detect the tracheal bifurcation.
- Demonstrated that mediastinum C.T. scans are more efficient than lung window C.T. scans for detecting the Carina, with faster processing times and fewer false positive Carina frames.
- Created a simple Windows app that lets the users draw rectangles around clear DBAs in the RLL region (only them now) to measure them for BAR and wall thickness.
- Updated the methods from our previous research [32] to measure the BAR in the ROIs of DBAs.
- Proposed a new method to measure airway wall thickness in the ROIs of DBAs.

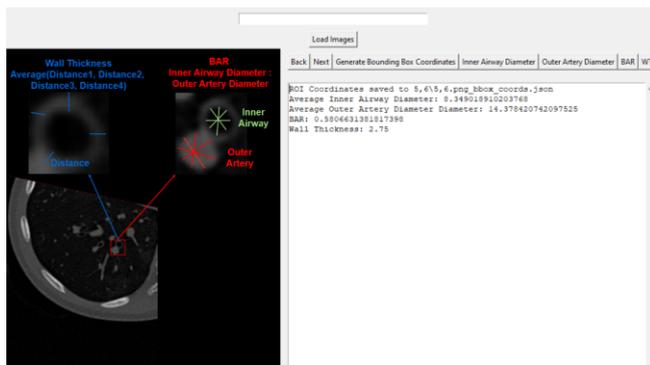

**Figure 1:** A simple app to measure inner airway and outer artery diameters, broncho-arterial ratio (BAR), and wall thickness.

## II. MATERIAL AND METHODS

### A. DATA DESCRIPTION

The original dataset includes C.T. scans from various scanners at Royal Darwin Hospital, Northern Territory, Australia, each with multiple resolutions and angles. Although CT scanners are primarily automated, settings often depend on the operator. HRCT scans require proper settings to ensure diagnostic quality, particularly for diagnosing airway diseases. This study utilized scans from a Philips Ingenuity Core 64 scanner, detailed in Table 1, which has scans of various slice thicknesses. It's part of a more extensive pediatric study on bronchiectasis at the same hospital, approved by the Menzies School of Health Research Ethics Committee (HREC-07/63, 22 April 2022).

**TABLE I**
TECHNICAL PARAMETERS: THE CATEGORIES AND THEIR ASSOCIATED FULL-LENGTH CT SCANS FROM DIFFERENT RECONSTRUCTION LENGTHS

| Scanner: Philips Ingenuity Core 64 | | | |
|---|---|---|---|
| Technical Parameters | | | |
| Acquisition Mode: Spiral (or Helical) | | | |
| Single Collimation Width: 0.625 mm | | | |
| Total Collimation Width: 64 x 0.625 = 40 mm | | | |
| Spiral Pitch Factor: 1.725; Kilovoltage Peak: 80 kVp | | | |
| Gantry Tilt: 0; DFOV (Average): 170 mm; Estimated Dose Saving (Average): -10 | | | |
| Slice Thickness | 0.67 mm | 1mm | 2mm |
| Full-length CT scans | 8 | 2 | 3 |

#### 1) Data Inclusion and Exclusion in Full-length CT Scan Collection

The subjects in the original collection had C.T. scans with slice thicknesses ranging from 0.67mm to 9mm, and not every subject had the same number of scans. We randomly selected 13 full-length C.T. scans of varying thicknesses (0.67mm, 1mm, 2mm) for our study to test our methods. We created two sets of images from the same scans to better visualize specific chest structures. One set of 13 scans is processed using "mediastinum window" settings, while the other (13) are processed with "lung window" settings, both utilizing the Sante DICOM Viewer supplied by our data providers. The mediastinum window is set to a window width (W.W.) of 300 HU (Hounsfield Units) and a window level (W.L.) of 50 H.U., enhancing soft tissue contrast and aiding in the identification of the tracheal bifurcation point where the Carina is located. The lung window is set with a W.W. of 1500 HU and a W.L. of -500 HU, optimizing the contrast between the air in the lungs and surrounding tissues, which is ideal for assessing lung parenchyma. These settings were applied across all scans to ensure uniform image quality.

#### 2) Implementation Resources

The image processing methods were implemented using Python 3.11.2 with Pillow, NumPy, SciPy, and Math libraries on a Windows 11 Pro computer equipped with a 12th Gen Intel Core i7-12700 processor (2.10 GHz) and 32 G.B. of RAM.

PyInstaller packages the Python scripts and Tkinter-based GUI components into a standalone executable, allowing the application to run on Windows P.C.s without requiring a Python installation. The associated program files, the application, and the instructions are available via a Google Drive, with a total size of approximately 1.6 GB.

https://drive.google.com/drive/folders/1HLgE0srIEwt6m94h3GCWrJ-_m0gdbGRP

### B. AUTOMATIC DETECTION OF THE RIGHT LOWER LOBE (RLL) REGION USING MARKERS

#### 1) CARINA DETECTION

The proposed method can be applied to full-length HRCT scans of different thicknesses because the RLL is present in all of them. Using Carina as a central marker, our method can approximately identify the starting point of the lower lobe region on the right side of the lung. Figures 2 & 6 provide the steps involved in detecting the Carina and extracting the RLL region. The RLL region is extracted (i.e., cropped) from all the frames until the end of the scan.

This has drastically reduced the processing time for locating and measuring the B.A. pairs for dilation and wall thickness.

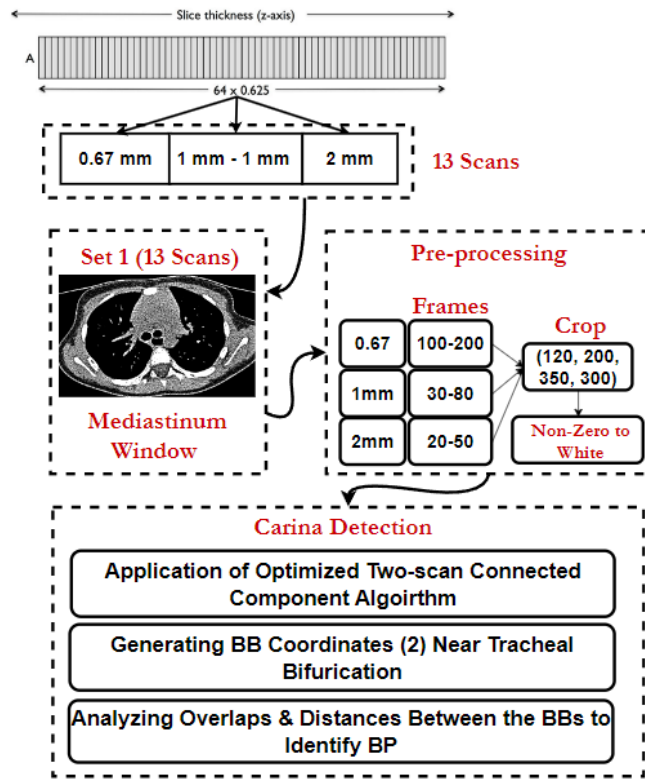

**Figure 2:** Steps for Detecting Approximate Tracheal Bifurcation Point

*Step 1: Preprocessing*
To locate the Carina in a full-length C.T. scan, it is unnecessary to analyze every frame or the entirety of each frame. Applying the proposed methods revealed that, for scans with 0.67mm thickness, the Carina is detected between the frames numbered 100 to 200; for 1mm-1mm scans, between frames 30 to 80; and for 2mm scans, frames 20 to 50. Within these specific frames, only the area within specific bounding box coordinates is used (cropped automatically), while the remaining areas are converted to black, as shown in Fig. 3. In all the scans, the bifurcation point is present within specific coordinates. These coordinates are adjusted based on the size of the frames in the C.T. scans by calculating and applying the scaling factors. For example, for frames of size 512x512, the bifurcation point and the surrounding region are present with the bounding box coordinates of (120, 200, 350, and 300), scaled if the size increases or decreases. The cropped regions are then processed to transform any non-zero pixels to white (255,255,255), as shown in Fig. 3. These modified frames are subsequently used further to determine the frame containing the bifurcation point. The visibility of the Carina on axial C.T. scans can vary based on the thickness of the slices, which affects the resolution of the images. Thinner slices, such as 0.67 mm, offer higher resolution, allowing for more precise identification of the Carina. On the other hand, thicker slices like 5 mm yield lower resolution, which may result in less detailed visualization of the Carina.

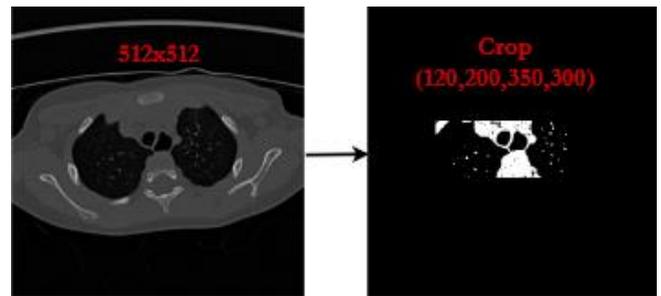

**Figure 3:** Original Frame (0.67 mm C.T. scan) (Left); Frame after applying the preprocessing methods (Right)

*Step 2: Carina Detection*
The algorithm employed in this study originates from our earlier publication [a], where it is described in detail. The current paper presents a concise version of the pseudocode (Pseudocode 1).

**Pseudocode 1**

1. Define connected_components(edges):
   1.1 Create dictionary for set of neighboring pixels
   1.2 Use DFS for finding connected nodes/pixels
   1.3 Generate components by exploring each unvisited node
2. Define make_edges(coordinates):
   2.1 Check eight neighboring pixels
   2.2 Yield edges if neighbors are in coordinates set
3. Define matching_pixels(image, is_black_enough):
   3.1 Yield pixel (0,0,0) from the image
4. Main Process:
   4.1 Get connected components from edges defined by matching black pixels.
   4.2 For each component:
      4.2.1 Calculate the bounding box and its area
      4.2.2 Store coordinates and draw a rectangle if the area is between 200 and 1000.

In an image, the 'connected_components' function finds the pixels that are connected to each other. Each component is a set of nodes connected directly or indirectly and not connected to any nodes outside the component. It uses depth-first search (DFS) to explore each node and its neighbors using a stack. It keeps track of visited nodes and records all nodes connected to the starting node as a component. In order to generate every possible pair of adjacent pixels using the provided coordinates, the 'make_edges' function is used. This is part of the first scan in the OCL algorithm. It checks each pixel's eight possible neighbors (8-connectivity) and yields a pair if the neighbor is also in the set of coordinates. This is used to build the graph of connected nodes. The 'matching_pixels' function generates coordinates of pixels in an image that meet a certain condition, which is used to find pixels of a specific intensity. The 'bounding box' function can use a collection of pixel coordinates to calculate the minimum and maximum X and Y coordinates. This allows for the creation of bounding boxes that will encompass all the specified pixels. Concurrently, the 'is_black_enough' function checks if a pixel's RGB values are entirely zeroes (0,0,0). Using the described functions (Pseudocode 1) and the area condition (greater than 200 and less than equal to 1500), bounding boxes are drawn on regions around the trachea and the large bronchi (main bronchi) where bifurcation occurs, as shown in Fig. 4. The coordinates for the bounding boxes are stored to be processed in further steps.

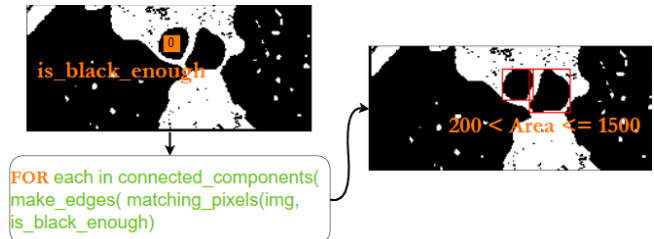

**Figure 4:** Using the OCL Algorithm and Area Condition to Draw Bounding Boxes around the trachea and the Main Bronchi

After filtering the bounding boxes based on a specific condition, the below algorithm (Pseudocode 2) checks for overlaps between the boxes in frames with two boxes. This is done by calculating the distance between them. The boxes are considered separate if the distance is more than three but less than or equal to 7. Here, the distance is calculated by measuring the horizontal (X-axis) distance between the leftmost X-coordinate ($x\_min_1$, $x\_max_1$) of the first bounding box and the rightmost X-coordinate ($x\_min_2$, $x\_max_2$) of the second bounding box. This is used to determine how far apart the two boxes are horizontally and identify the bifurcation point by selecting the lowest distance, also known as the Carina, which is shown in Figure 5.

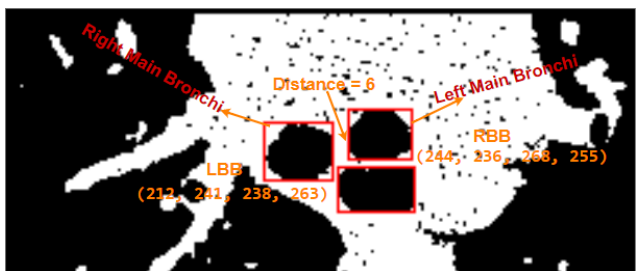

**Figure 5:** Checking Overlaps and Calculating Distance to Identify the Bifurcation Point

---

**Pseudocode – 2**
1. Define check_overlap(box1, box2):
   1.1 Extract the *min* and *max* X and Y coordinates from both boxes
2. Define calculate_distance(box1, box2):
   2.1 Extract the *min* and *max* X coordinates from both boxes
   2.2 Calculate the horizontal distance between the closest edges of the two boxes
3. Define process_bounding_boxes(file path):
   3.1 Load Bounding Box Coordinates from a CSV file
   3.2 For each entry:
       3.1.1 Check for non-overlap and calculate distance
       3.1.2 If the distance is between 5 and 7, append the results

---

### 2) RIGHT LOWER LOBE (RLL) REGION EXTRACTION

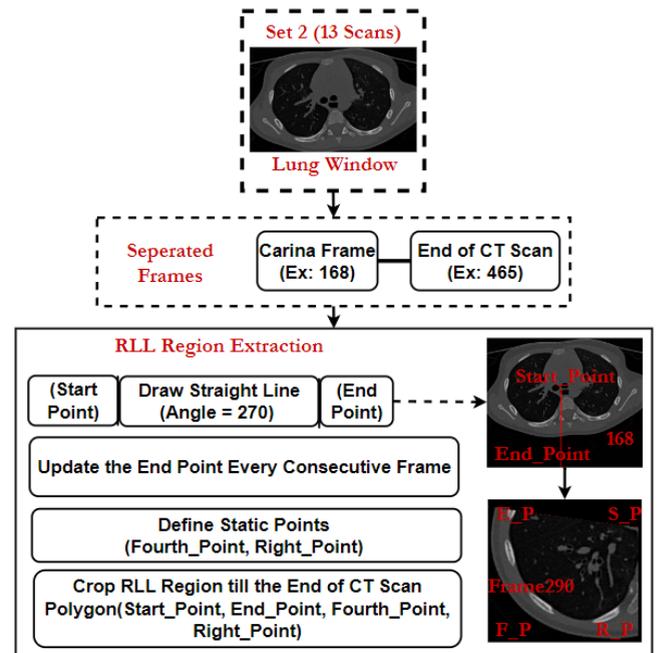

**Figure 6:** Right Lower Lobe Region Extraction Process from the Lung Window CT scans

The right lower lobe of the lung typically begins below the bifurcation point of the trachea. After identifying the approximate bifurcation point (B.P.), a straight line is drawn downward and leftward from B.P., which serves as the starting point. This point is determined by taking the greater x-coordinate and the smaller y-coordinate from two bounding boxes, ensuring it starts at the rightmost and lowest corner. The line extends toward the image boundary for the endpoint, calculated dynamically based on the image dimensions and a specified angle of 270. This calculates the maximum distances to the bottom edges of the image from the starting point. These distances are adjusted by dividing the cosine and negative sine of the angle in radians, respectively, to account for the direction and stretch caused by the angle. The minimum of these adjusted distances determines the length of the line, ensuring it does not exceed the image frame. This process is described in Pseudocode 3 and illustrated in Fig. 6 & 7.

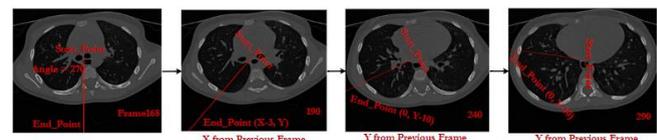

**Figure 7:** Start_Point and End_Points Across Various Farmes in the Scan after the Carina Frame

```
                        Pseudocode – 3
BEGIN
    IF angle_degrees IS BETWEEN 180 and 360
        SET max_x_dist TO -start_point[0]   //Distance to the left edge
        SET max_y_dist TO img_height – start_point[1]
                                        //Distance to the bottom edge
    SET max_x_dist TO max_x_dist / cosine(angle_radians)
    SET max_y_dist TO max_y_dist / -sine(angle_radians)
                                        //Adjust distance based on angle

    SET length TO minimum OF [max_x_dist] AND [max_y_dist]

    SET endpoint_x TO start_point[0] + (length * cosine(angle_radians)
    SET endpoint_y TO start_point[1] – (length * sine(angle_radians)

    SET endpoint TO (integer of endpoint_x, integer of endpoint_y)

    RETURN endpoint
END
```

The endpoint coordinates are then calculated by adjusting the starting point coordinates by this length, factoring in the cosine and sine of the angle. The X-coordinate is derived by adding the cosine of the angle, multiplied by the line's length, to the x-coordinate of the starting point, effectively moving the line horizontally. Concurrently, the Y-coordinate is calculated by subtracting the sine of the angle multiplied by the line's length from the y-coordinate of the starting point. This subtraction accounts for the fact that the y-axis values increase in image coordinates as one moves downward.

The next sub-step dynamically updates and draws a line across sequential C.T. scan frames after the Carina frame. This is done by adjusting the line's endpoints in each frame and logging these changes. Starting with coordinates (X) and (Y) calculated from the Carina frame (approx.), the X coordinate is decremented by ten units (X-10) in each subsequent frame until X reaches 0, resulting in a leftward movement along the X-axis. Once X reaches 0, the Y coordinate is decreased by three units (Y-3) in the subsequent frames, resulting in an upward movement along the Y-axis.

In the next sub-step, two static points, 'right_point' and 'fourth_point', are defined to complete a polygon with four points (start_point, end_point, fourth_point, right_point). The 'right_point' is set at the X-coordinate of the 'start_point' and a fixed Y-coordinate of 511. The 'fourth_point' is set at the left edge of the image (X-coordinate of 0) and the same Y-coordinate of 511. Using these points, a polygon is drawn in each frame to isolate the respective region (crop). These regions in each frame are approximately highlighting the right lower lobe of that frame, as shown in Fig. 8.

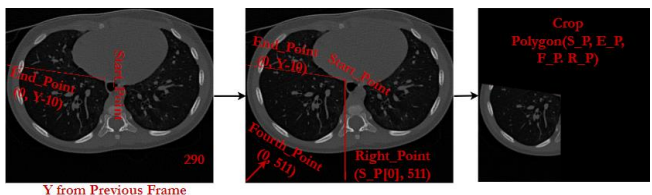

**Figure 8:** Defining Two Static Points (F_P & R_P) to Crop the Polygon Region that is RLL

## C. MEASURING DISCRETE BA PAIRS IN THE RLL REGION FOR BRONCHO-ARTERIAL RATIO (BAR) AND WALL THICKNESS

The BA pairs in an HRCT scan can appear in various sizes, shapes (irregular), and forms with different intensities (RGB). They can be large, medium, or small and appear in discrete pairs or accompanied by pulmonary veins, depending on the broncho-pulmonary segment and lobe they are situated in (Fig. 1A). Similarly, the RGB intensities of the airway walls can also differ based on their location (Fig. 1B). These are some of the essential aspects to keep in mind while developing methods for detecting discrete BA pairs in HRCT scans. In our previous study, we proposed methods to automatically locate all the objects and detect discrete BA pairs in all the frames of a full-length C.T. scan [32], where the methods are explained in detail. In this study, our aim is not to locate all the objects in the frames but to locate the objects inside the discrete B.A. pairs circled by a human and measure them for broncho-arterial ratio (BAR) and wall thickness. This new approach is operator (radiologist, physician, etc.) dependent, as they need to circle (ROI) the B.A. pairs they need to measure.

### 1) MEASURING BRONCHO-ARTERIAL RATIO (BAR)
The process involves five steps to measure the BAR by processing the region of interest (ROI).
- Extracting ROIs of discrete BA pairs.
- Processing/Enhancing ROIs: Preprocess gray-scaled ROI patches by replacing pixel values to enhance regions. For airways, pixels with R-value <= 25 are turned to black (0,0,0) and others to gray (80,80,80). Similarly, the conditions rgb[0] <= 25, 25 < rgb[0] <= 45, and 45 < rgb[0] <= 100, are used for the artery region to replace the pixel values with (0,0,0), (255,0,0) and (80,80,80) respectively (Fig. 12). This creates two processed ROIs.
- Detecting Matching Sequences and Constructing Coordinates: This step's algorithm [32] checks for 'matched sequences' in the pixel values of each row of a Processed ROI and constructs the coordinates for the inner airway and artery regions. The sequences represent a pattern to detect the respective regions. The sequences Size1 through Size13, like [80,0,80] to [80,0,…,0,80], detect airway regions. The '80's at each end represents airway walls, and the increasing zeros in between indicate airway spaces. For arteries, sequences Size7 through Size22, from [255,80,…,80,255] to [255,80,…,80,255], start and end with '255', marking arterial boundaries, while the '80s' in the middle suggest arterial passages. These patterns help detect changes in airway and arterial regions, which are helpful in analyzing abnormalities.

For example, in **Row N-1** of Fig. 9, the list of pixels in the processed artery ROI as [0, 255, 80, 80, 80, 80, 80, 80, 80, ……, 255, 0, 0, …., 0]. The algorithm then searches for matching sequences of pixel values and notes their positions in a range format. For instance, if sequences are detected at row 465 in the image, the resulting ranges would be (254, 270) & (506, 524), which indicates the columns in the image. This method is applied to all rows, gathering coordinates for the inner airway and outer artery areas. Typically, these coordinates accurately outline the intended areas in the processed images of the airway and artery. Occasionally, there may be instances when some coordinates might be missing. These missing parts can be added later by using a (2,2) kernel.

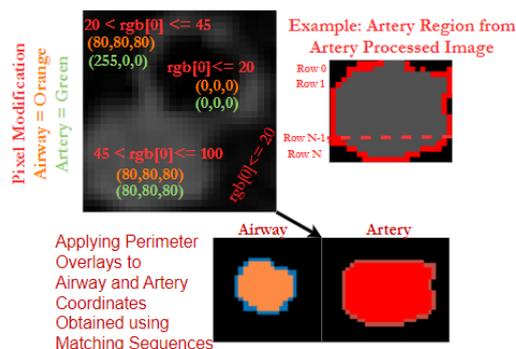

**Figure 9:** Proc. Conditions, Perimeter Overlays for Inner Airway and Outer Arteries

- Obtaining Perimeter Coordinates and Calculating Euclidean Distances: We applied traditional methods such as detecting contours or blobs by filling the constructed coordinates with different pixel values. These techniques are helpful in determining the coordinates of the object contours (inner airway, artery) within the filled regions, as shown in Figure 9. They are outlined using the draw contours () function in OpenCV, which overlays the contours onto the ROI with a thickness of 1, highlighting the contour boundaries or perimeters as illustrated in Fig. 9. Using these coordinates, Euclidean distances are calculated between each pair of boundary pixels of the objects (Fig. 10).
- Diameter Selection and BAR Computation: After calculating all the distances for the combinations, they are appended to a list in the following format: (start point $(X_1, Y_1)$, endpoint $(X_2, Y_2)$, Distance). The Euclidean distance is measured in pixels (px). Using a lambda function, the combination with the 'maximum Euclidean distance' (MED) is also a measure of an object's length. The MED, or major axis, is the longest line that can fit within the item and typically passes through or near the object's center.

Using the major axis, the algorithm [32] for this step determines additional diameters, either four or six, depending on the object's size. If the length exceeds 20 px, six diameters are computed; if it is shorter, four are calculated. The process involves adjusting the coordinates of the major axis at both ends—retracting one end by -4 pixels and extending the other by +4 pixels. This modification creates two additional endpoints, which usually go through or near the object's center, as shown in Figure 10. This step is repeated until the required diameters (including the major axis and others) are determined. Finally, the algorithm calculates the Euclidean distance between these endpoints and averages them to choose the diameters of the inner airways and outer arteries, as illustrated in Fig. 10.

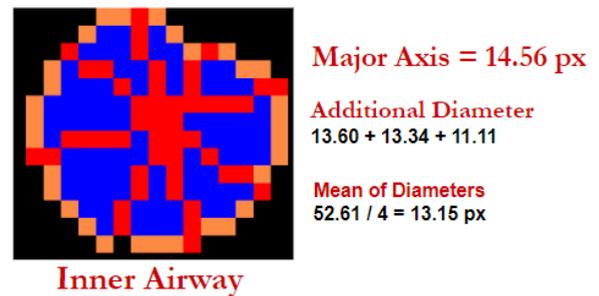

**Figure 10:** Process for Obtaining Diameter for Inner Airway Region (Example)

*2) MEASURING BRONCHIAL-WALL THICKNESS*

Bronchiectasis is a condition where the bronchi abnormally widen, often making the assessment of airway wall thickness crucial for its detection. Measuring the wall thickness typically involves identifying the inner and outer boundaries of the airway walls. The thickness of the walls can be determined by measuring the Euclidean distance between these boundaries. Due to irregular geometries, measuring wall thickness on elliptical and non-circular objects can be a bit more complex than measuring uniform shapes. Generally, the wall thickness tends to decrease with each subsequent generation because the diameter of the airways decreases. However, airway pressure and specific mechanical characteristics of the wall, such as the presence of cartilage, muscle, and connective tissue, also change and influence the overall thickness.

To calculate the BWT, we extracted the perimeter coordinates for the inner and outer airway regions. Then, the Euclidean distance is calculated between the perimeter coordinates of both boundaries at multiple points (4) and averaged to get the final wall thickness. The following steps explain the process involved in measuring the wall thickness:

- Process the ROI by turning pixels with an R-value ≤ 20 to black (0,0,0) and others to gray (80,80,80).
- Identify specific gray pixels (r=80, g=80, b=80) and record their coordinates.
- Load both inner airway perimeter (IAP) coordinates and gray coordinates, setting a distance threshold of 4 to determine proximity. This helps in precise anatomical mapping.
- Identify gray coordinates within the specified distance from any IAP coordinate to define the outer airway region.

- Determine edge coordinates from the gray coordinates within the distance threshold, forming the outer airway perimeter.
- Select random coordinates from each cardinal direction (north, south, east, west) from the inner airway edge coordinates for representative analysis, ensuring new random coordinates each time.
- Find the farthest points in each direction from the outer airway edge coordinates based on the initial cardinal coordinates.
- Calculate the Euclidean distance between corresponding points (north, south, east, west) from inner and outer airway boundaries and compute their average to determine wall thickness.

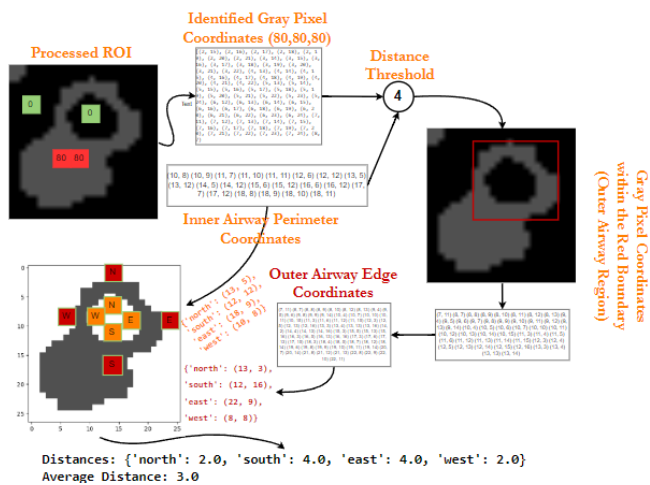

**Figure 11:** Steps for measuring bronchial-wall thickness

## III. RESULTS & DISCUSSION

### A. CARINA DETECTION

To detect the approximate frame containing the bifurcation point (B.P.) (i.e., Carina), we have used the Mediastinum window C.T. scans (Set 1). The mediastinum window makes it easier to identify the Carina. The clarity and detail required to detect the Carina can be obscured in the lung window due to its focus on-air content. This often leads to a loss of detail in soft tissue structures, crucial for accurately locating the Carina.

TABLE I
**F_I**: Frames Identified, **BB_C**: Bounding Box Coordinates, **D**: Distance, **C_F**: Carina Frame, **S_1, S_2, S_3**: Time Taken To Run Step 1, Step 2, Step 3 In Seconds

| CT Scan | F_I, BB_C (2) & D | C_F | S_1 | S_2 | S_3 |
|---|---|---|---|---|---|
| **0.67mm** | **Between Frame 120 - 200** | | | | |
| 465 512x512 | 169, [(212,241,238,263), (244,236,268,255)] & 6 | 169 | 1.20 | 320.39 | 0.012 |
| | 170, [(212,241,238,263), (244,236,268,256)] & 6 | | | | |
| | + 2 more | | | | |
| 420 512x512 | 129, [(226,214,250,227), (202,215,220,232)] & 5 | 129 | .86 | 326.38 | 0.01 |
| | 131, [(255,219,314,236), (217,226,249,252)] & 4 | | | | |
| | +8 more | | | | |
| 537 512x512 | 151, [(213,227,245,252), (249,218,277,237)] & 4 | 151 | 1.54 | 332.97 | 0.01 |
| | 152, [(212,227,245,253), (250,218,278,237)] & 5 | | | | |
| | +2 more | | | | |
| 535 512x512 | 172, [(200,267,225,286), (231,265,255,278)] & 6 | 172 | 1.47 | 328.46 | 0.01 |
| | 174, [(198,268,227,286), (233,265,257,279)] & 6 | | | | |
| | +1 more | | | | |
| 535 512x512 | 159, [(209,222,236,242), (240,238,266,254)] & 4 | 159 | 1.37 | 330.75 | 0.01 |
| | 160, [(241,215,268,232), (208,223,236,244)] & 5 | | | | |
| | +7 more | | | | |
| 553 512x512 | 171, [(207,277,228,294), (233,271,249,285)] & 5 | 171 | 1.02 | 345.80 | 0.01 |
| | 172, [(206,276,228,294), (234,271,250,285)] & 6 | | | | |
| | +1 more | | | | |
| 402 512x512 | 145, [(212,216,232,233), (237,212,257,226)] & 5 | 145 | .98 | 338.20 | 0.01 |
| | 146, [(211,216,231,232), (238,212,258,225)] & 7 | | | | |
| | +7 more | | | | |
| 465 512x512 | 119, [(239,218,258,236), (265,223,285,234)] & 7 | 119 | .98 | 327.76 | 0.01 |
| | 120, [(238,217,258,235), (265,223,284,234)] & 7 | | | | |
| | +15 more | | | | |
| **1mm** | **Between Frame 30 - 80** | | | | |
| 180 | 51, [(225,238,257,264), (261,230,289,249)] & 4 | 51 | .67 | 185.87 | 0.005 |
| | +4 more | | | | |
| 336 | 55, [(240,272,265,286), (204,275,230,294)] & 10 | 55 | .58 | 178.25 | 0.005 |
| **2mm** | **Between Frame 10 – 40** | | | | |
| 90 536x536 | 28, [(201, 248, 245, 269), (251, 252, 284, 272)] & 6 | 28 | .35 | 108.90 | 0.003 |
| | + 2 more | | | | |
| 89 552x552 | 31, [(256, 300, 278, 312), (284, 280, 309, 291)] & 6 | 31 | .35 | 112.6 | 0.004 |
| 66 440x440 | 25, [(192, 257, 212, 270), (205, 240, 221, 250)] & 7 | 25 | .54 | 124.41 | 0.006 |
| | +4 more | | | | |

The proposed method for identifying Carina's approximate bifurcation point is applied to mediastinum window C.T. scans of 0.67mm and 1mm thicknesses. For 2mm scans, we used lung window C.T. scans, which are explained in detail later (Fig. 13). The results and the time required for these processes are detailed in Table I. For 0.67mm scans, the carina frame (C_F) typically appears between frame numbers 120 and 200. For 1mm and 2mm scans, the C_F is detected between frame numbers 30-80 and 10-40, respectively. This method avoids processing all the frames to reduce unnecessary analysis time. In all 0.67mm and 1mm scans with size of 512x512, the bifurcation point lies within the bounding box coordinates (120,200,350,300). For different frame sizes, these coordinates are appropriately scaled, which is a straightforward adjustment. For 2mm scans with a frame size of 552x552, the coordinates adjust to (129,216,377,324) for effective cropping. The step labeled S_1 indicates the time it takes to preprocess a full-length C.T. scan, including frame separation, cropping, and pixel enhancement, typically around or less than one second, as shown in Table 1. In some cases where the C.T. scan exceeds 500 frames, the visibility of the branching into the right and left main bronchi may be poor or incomplete, as illustrated in Figure 12, posing challenges in creating accurate bounding boxes. To address this, an additional preprocessing step involving dilation using a 3x3 kernel has been introduced to improve the clarity of these frames.

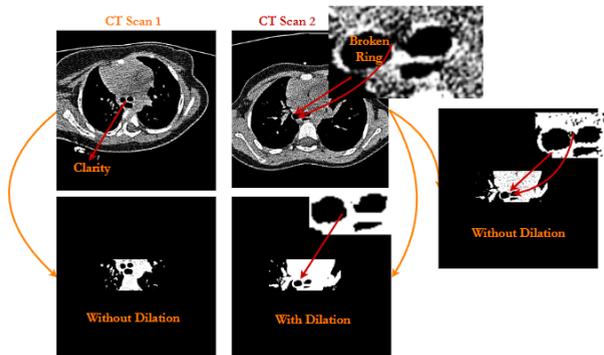

**Figure 12:** Applying dilation to improve visibility of the bifurcation point

The S_2 indicates the duration needed to implement the optimized component labeling algorithm [32] to outline bounding boxes around the right and left main bronchi in multiple separated and cropped frames. For 0.67mm scans, this step requires less than 6 minutes (360s) as it processes just 80 frames with cropped regions from over 400 in total. Similarly, for 1mm and 2mm scans, the time is about 3 minutes (~180s), covering 50 and 30 frames, respectively, from their full-length scans. The S-3 measures the time to analyze overlaps and calculate distances between the bounding boxes (S_2) to locate the bifurcation point (i.e., Carina), typically taking less than a second in most scans. The second column in Table. 1 showcases the number of frames potentially containing the Carina, along with their respective bounding box coordinates and the distances measured between the right and left main bronchi as identified by the proposed method. On average, the total time to execute all three steps and determine the approximate Carina frame is under 6 minutes (360s) for 0.67mm scans, about 3 minutes (180s) for 1mm scans, and around 2 minutes (120s) for 2mm scans.

In some mediastinum C.T. scans, particularly those with a 2mm thickness, the bifurcation point is too rapid and indistinct to be seen clearly (Fig. 13), making it challenging to identify using our methods. We have utilized lung window C.T. scans to detect the Carina in such cases. As the slice thickness increases beyond 2mm, it becomes increasingly difficult for the proposed methods to accurately identify the Carina. This issue may arise from the reduced resolution in the thicker slices, where smaller structures like the Carina may not appear as distinct. Additionally, the partial volume effect can make it tough to accurately outline the edges of structures such as the Carina.

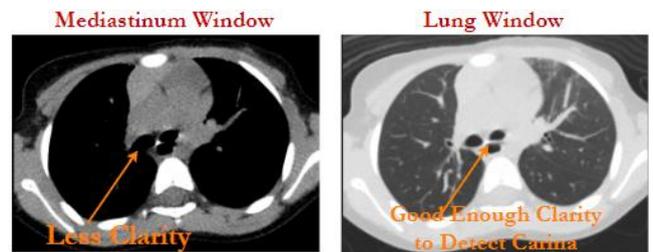

**Figure 13:** Good Clarity of bifurcation point in Lung window C.T. scan (right) than Mediastinum window (left) in 2mm scans.

The proposed method can also identify the Carina in lung window C.T. scans across various thicknesses. The main difference between mediastinum and lung window C.T. scans is in the processing times and the number of potential Carina frames detected. A comparative analysis of the two is outlined in Table II, utilizing the first three scans in Table I. In lung window C.T. scans with a 0.67mm thickness, there is a significant increase in both the number of potential Carina frames and the processing times compared to other window scans, as shown in Table II. This increase is attributed to the OCL algorithm generating false positive bounding boxes in the lung window scans caused by unnecessary pixel data in the frames.

**TABLE II**
COMPARISON BETWEEN MEDIASTINUM AND LUNG WINDOW CT SCANS
**T_P_CF:** TOTAL POTENTIAL CARINA FRAMES IDENTIFIED; **C_F & D:** CARINA FRAME & DISTANCE BETWEEN BOUNDING BOXES; **S_1, S_2, S_3:** TIME TAKEN TO RUN STEP 1, STEP 2, STEP 3 IN SECONDS

| M_W | | | L_W | | |
|---|---|---|---|---|---|
| T_P_CF | C_F & D | S_1 + S_2 + S_3 | T_P_CF | C_F & D | S_1 + S_2 + S_3 |
| 4 | 169 & 6 | 1.20 + 320.39 + 0.01 | 13 | 167 & 4 | 1.98 + 560 + 0.01 |
| 10 | 129 & 5 | .86 + 326.38 + 0.01 | 81 | - | 2.82 + 534.8 + 0.02 |
| 4 | 151 & 4 | 1.54 + 332.97 + 0.01 | 23 | 148 & 5 | 2.94 + 601.2 + 0.01 |

It should be noted that the Carina is not identified in some scans, as illustrated in Fig. 14. This may be attributed to differences in growth patterns, development factors, and possible pathological conditions.

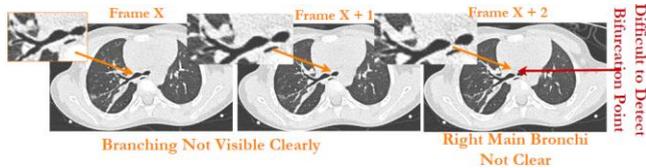

**Figure 14:** Difficulty in detecting bifurcation point due to unclear branching

### B. RLL REGION EXTRACTION

Once the approximate Carina Frame (C_F) is identified in a full-length C.T. scan, only the frames following the C_F are selected for further analysis. For instance, in a 0.67mm scan with 465 frames, if the C_F is at 169, then 297 subsequent frames are processed for Right Lower Lobe (RLL) region extraction. This extraction utilizes lung window C.T. scans (Set 2) because the main goal is to detect and assess broncho-arterial pairs for indications of bronchial dilatation and wall thickness, which are not discernible in mediastinum window scans. The total frames (T_F) listed in the first column of Table III are derived from the complete lung window C.T. scans of different thicknesses. The RLL typically begins just below the bifurcation point marked by the Carina frame. The first step involves drawing a straight line from the bifurcation point at a 270-degree angle toward the frame's end to establish initial endpoint coordinates (x, y), taking less than 0.1 seconds as this is applied only on the Carina frame. Subsequently, step 2 adjusts the endpoint coordinates in later frames based on specific conditions. For 0.67mm scans, the X coordinate decreases by ten units until it hits zero, after which five units reduce the Y coordinate until the scan ends. For 1mm and 2mm scans, the updating rules are (X – 25, Y) followed by (0, Y -12), and (X – 50) followed by (0, Y – 25), respectively. The processing time for this method ranges from about 15 to 25 seconds for 0.67mm scans and less than 10 and 5 seconds for 1mm and 2mm scans, respectively.

TABLE III
RLL EXTRACTION PROCESSING TIMES
T_F: TOTAL FRAMES

| T_F from Carina to End of Scan | Step 1 (S) | Step 2 (S) | Step 3 (S) |
|---|---|---|---|
| **0.67mm** | E_P - (X – 10, Y) and then (0, Y – 5) | | |
| 297 | <0.1 | 15.19 | 8.07 |
| 288 | <0.1 | 16.25 | 7.67 |
| 387 | <0.1 | 25.87 | 14.78 |
| 364 | <0.1 | 22.23 | 9.79 |
| 377 | <0.1 | 22.20 | 10.38 |
| 383 | <0.1 | 23.80 | 9.71 |
| 258 | <0.1 | 15.76 | 6.26 |
| 347 | <0.1 | 21.31 | 9.66 |
| **1mm** | E_P - (X – 25, Y) and then (0, Y – 12) | | |
| 130 | <0 | 8.31 | 3.10 |
| 114 | <0 | 6.3 | 2.45 |
| **2mm** | E_P - (X – 50, Y) and then (0, Y – 25) | | |
| 63 | <0 | 3.43 | 1.45 |
| 59 | <0 | 3.19 | 2.36 |
| 42 | <0 | 2.85 | 1.52 |

Step 2 enables the techniques to approximately trace the oblique fissure within the C.T. scan, a key indicator for locating the right lower lobe region, as depicted in Fig. 18. By using a fixed start point and dynamic endpoints to draw a straight line across all frames, the oblique fissure is then manually reviewed to assess the effectiveness of the proposed methods (Fig. 15). In most instances, the drawn line appears either slightly above or at an angle to the oblique fissure, as illustrated in Fig. 15, confirming that our method successfully identifies the right lower lobe region. Even if the line falls significantly above the oblique fissure, it's not a significant concern, as our primary goal is to isolate the RLL or the targeted area from the rest of the frame. This separation helps enhance processing speeds and facilitates more effective analysis of the different elements within that region. In step 3, two additional static coordinates are defined: the fourth and right points. In each frame, these three static points (start point, fourth point, right point) and one dynamic point (endpoint) are used to form a polygon. This polygon allows for the cropping of the internal area, which roughly corresponds to that frame's right lower lobe region. The implementation of this process takes about 5 to 10 seconds for 0.67mm scans and less than 5 seconds for 1mm and 2mm scans. The cropped areas are further processed to identify the B.A. pairs for measurements of broncho-arterial ratio (BAR) and wall thickness.

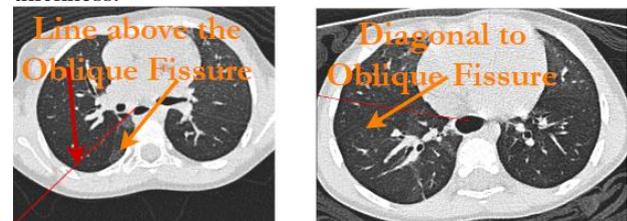

Figure 15: The Red line (automated) appearing above (Left) and diagonal to the oblique fissure (right)

### C. MEASURING BRONCHO-ARTERIAL RATIO (BAR) AND WALL THICKNESS IN THE REGION OF INTEREST (ROI)

We have developed a straightforward interface that allows the detection of bronchi-arterial (B.A.) pairs. Our methods then automatically provide measurements for the inner airway diameter, outer airway diameter, and outer artery diameter, enabling effective assessment of broncho-arterial ratios and wall thicknesses. A link to the interface is provided in the Resources section (Section. II.A.2). This can aid us in the future to better understand various conditions related to airways and vascular structures by enabling quicker analysis compared to manual methods. These conditions include bronchiectasis, chronic obstructive pulmonary disease (COPD), and asthma. In this study, ten discrete B.A. pairs (DBA) pairs are circled using the tool to apply the methods (Section II.C) and measure them for BAR and wall thickness. The measurements are validated by an expert on our team, allowing us to update the methods based on their assessment. The measurements from the proposed methods have a deviation of ±0.10mm - 0.15mm from a clinician's measurements. In most cases, the clinician measured the bronchial-wall thickness by dividing the difference between outer airway and inner airway diameters by 2. In this research, we proposed a novel method to measure wall thickness, which yielded similar results. Validations from multiple clinicians are required to address the intra and inter-rater variability and improve the methods even further. Table. IV & V provides a comparison between the proposed methods and expert readings. The research aims to develop 2D methods that can be expanded in the future to perform 3D reconstruction of various structures and facilitate more comprehensive analysis and measurement of broncho-arterial abnormalities.

TABLE IV
MEASUREMENTS FROM THE PROPOSED METHODS
**DBAP:** DISCRETE BA PAIRS; **PS:** PIXEL SPACING; **IAD:** INNER AIRWAY DIAMETER; **ARD:** OUTER ARTERY DIAMETER; **BAR:** BRONCHO-ARTERIAL RATIO; **WT:** WALL THICKNESS

| DBAP | IAD | ARD | BAR | WT |
|---|---|---|---|---|
| 1 | 1.81 | 2.77 | .65 | 0.8 |
| 2 | 2.58 | 3.19 | .81 | 0.54 |
| 3 | 3.45 | 4.78 | .72 | 1 |
| 4 | 2.43 | 2.59 | .94 | .71 |
| 5 | 2.63 | 3.35 | .79 | .82 |
| 6 | 3.51 | 3.72 | .94 | .82 |
| 7 | 2.64 | 3.42 | .77 | .91 |
| 8 | 2.91 | 4.11 | .71 | .91 |
| 9 | 2.66 | 3.36 | .79 | .73 |
| 10 | 3.01 | 3.90 | .81 | 0.82 |

TABLE V
MEASUREMENTS FROM AN EXPERT
**EIAD:** EXPERT INNER AIRWAY DIAMETER; **EOAD:** EXPERT OUTER AIRWAY DIAMETER; **EARD:** EXPERT ARTERY DI; **E.B.:** EXPERT BAR; **E.W.:** EXPERT WALL THICKNESS

| DBAP | EIAD | EOAD | EARD | EB | E.W. |
|---|---|---|---|---|---|
| 1 | 1.9 | 3.1 | 2.60 | .73 | 0.7 |
| 2 | 2.7 | 3.8 | 3.42 | .79 | 0.5 |
| 3 | 3.4 | 5.4 | 4.72 | .72 | 1 |
| 4 | 2.5 | 3.9 | 2.84 | .88 | 0.7 |
| 5 | 2.7 | 4.3 | 3.10 | .87 | 0.8 |
| 6 | 3.4 | 4.7 | 3.82 | .89 | 0.65 |
| 7 | 2.7 | 4.6 | 2.62 | 1.03 | 0.95 |
| 8 | 3.1 | 4.9 | 3.97 | 0.78 | 0.9 |
| 9 | 2.7 | 4.1 | 3.51 | 0.77 | 0.7 |
| 10 | 3.2 | 4.9 | 3.90 | 0.82 | 0.85 |

## IV. CONCLUSION

Our study introduces a novel approach that differs from existing methods. Our primary focus is on developing techniques to detect and measure structures within the right lower lobe (RLL) of the lung and introducing three novel methods. Given the variations in orientation, angled nature, and partial volume effect in the upper and middle lobes, we argue that each lobe requires specific image processing and measurement methods to ensure accurate results. For the first method, 13 full-length HRCT scans of various thicknesses (0.67mm, 1mm, 2mm) were used to apply the methods and identify the tracheal bifurcation point, isolating the frames containing the RLL region. Results showcased accurate detection of the tracheal bifurcation frame in most cases, with a deviation of ±2 frames. To implement the second and third methods, a Python-based Windows application is created to draw bounding boxes around ten broncho-arterial (B.A.) pairs and apply the proposed methods to measure inner airway and artery diameters, B.A. ratio (BAR), and wall thickness. A human reader then validated these measurements, demonstrating comparable accuracy.

Our current paper lays the groundwork for future research into 3D imaging methods that could improve the accuracy, precision, and reliability of assessing and monitoring changes in the bronchial tree, ultimately leading to improved diagnosis and treatment plans.

multi-domain airway tree modeling. *Medical Image Analysis*, 90, p.102957.
52. LEBA, C. and SEWELL, J., 2023. Evaluation of a 3D-printed, color-coded tracheobronchial tree for bronchoscopy training and anatomy learning. *Chest*, 164(4), p.A3801.
53. Bauer, C., Eberlein, M. and Beichel, R.R., 2014. Graph-based airway tree reconstruction from chest C.T. scans: evaluation of different features on five cohorts. *IEEE transactions on medical imaging*, 34(5), pp.1063-1076.
54. Bauer, C., Krueger, M.A., Lamm, W.J., Smith, B.J., Glenny, R.W. and Beichel, R.R., 2013. Airway tree segmentation in serial block-face cryomicrotome images of rat lungs. *IEEE transactions on biomedical engineering*, 61(1), pp.119-130.
55. Tan, W., Yang, J., Zhao, D., Ma, S., Qu, L. and Wang, J., 2012, May. A novel method for automated segmentation of airway tree. In *2012 24th Chinese Control and Decision Conference (CCDC)* (pp. 976-979). IEEE.
56. Liu, X., Chen, D.Z., Tawhai, M.H., Wu, X., Hoffman, E.A. and Sonka, M., 2012. Optimal graph search based segmentation of airway tree double surfaces across bifurcations. *IEEE transactions on medical imaging*, 32(3), pp.493-510.
57. DeGrave, A.J., Janizek, J.D. and Lee, S.I., 2021. A.I. for radiographic COVID-19 detection selects shortcuts over signal. Nature Machine Intelligence, 3(7), pp.610-619.
58. Park, S.H. and Han, K., 2018. Methodologic guide for evaluating clinical performance and effect of artificial intelligence technology for medical diagnosis and prediction. *Radiology*, 286(3), pp.800-809.
59. Kelly, C.J., Karthikesalingam, A., Suleyman, M., Corrado, G. and King, D., 2019. Key challenges for delivering clinical impact with artificial intelligence. BMC medicine, 17, pp.1-9.
60. Geirhos, R., Jacobsen, J.H., Michaelis, C., Zemel, R., Brendel, W., Bethge, M. and Wichmann, F.A., 2020. Shortcut learning in deep neural networks. Nature Machine Intelligence, 2(11), pp.665-673.
61. Towards trustable machine learning. Nat Biomed Eng 2, 709–710 (2018). https://doi.org/10.1038/s41551-018-0315-x.

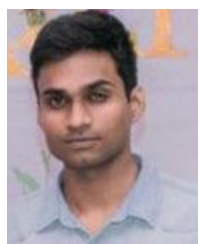

**ABHIJITH REDDY BEERAVOLU** is pursuing a Post-Doctoral Biomedical Engineering degree at Charles Darwin University, Casuarina, NT, Australia. His thesis is on "providing interpretability for diverse medical A.I. systems." This requires working with real-world medical data encompassing images, videos, clinical records, and text to create diagnostic tools and model-specific interpretability techniques. He aims to improve local and global communities by generating innovative ideas to address real-world challenges.

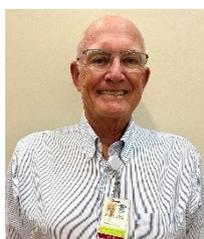

**Dr. BRENT MASTERS** has been a pediatric respiratory physician since 1984 in Calgary, Canada, and Brisbane, Australia. He completed his Thoracic medicine training in Melbourne and gained additional experience in Calgary. Serving as the head or director of pediatric respiratory medicine at major hospitals in Brisbane since 1987, Dr. Masters has extensive clinical expertise, including outreach work in Aboriginal and Torres Strait Island communities. His specializations include flexible bronchoscopy, respiratory teaching, and integrating C.T. and bronchoscopy 3D anatomy. He has trained professionals from nine countries, conducted workshops, and contributed to over 150 peer-reviewed publications. Additionally, he has served as an advisor and committee member for the TSANZ (Thoracic Society of Australia and New Zealand) and RACP (Royal Australasian College of Physicians) on matters related to pediatric bronchoscopy training.

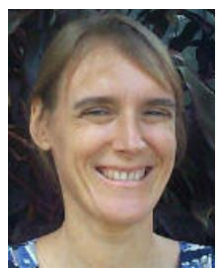

**MIRJAM JONKMAN** is currently a Lecturer and a Researcher with the Faculty of Science and Technology, Charles Darwin University, Casuarina, NT, Australia. Her research interests include biomedical engineering, signal processing, and the application of computer science to real-life problems.

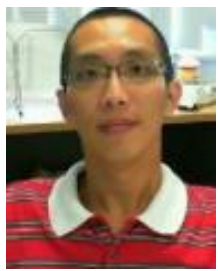

**Dr. KHENG CHER YEO** is currently a Senior Lecturer in information technology with the College of Engineering, IT and Environment, Australia. He is passionate about teaching and has taught hardware, mathematics, networking, software engineering, and project management. He is also active in research and his research interests include the areas of intelligent signal processing and control, networking, and security and app development

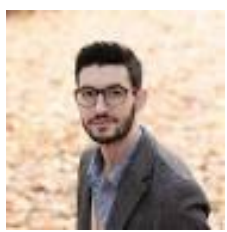

**Dr. SPYRIDON PROUNTZOS** is a Radiologist and Ph.D. candidate in Pediatric Radiology at the National and Kapodistrian University of Athens, Athens, Greece, and works at Attikon University General Hospital. He has a strong interest in Body Imaging and Pediatric Radiology, with expertise in Magnetic Resonance, Computed Tomography, Medical Imaging, and Diagnosis.

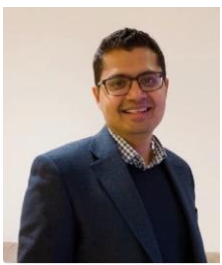

**Dr. RAHUL J THOMAS** is a respiratory and sleep pediatrician. He is an NHMRC PhD scholar and supported by the Queensland Hospital Foundation and the CRE in his field of interest, i.e., large airway diseases and diagnostic modalities of flexible bronchoscopy and thoracic radiology.

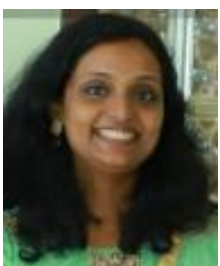

**Dr. EVA IGNATIOUS** is currently a Ph.D. Researcher with Charles Darwin University, Australia. Her research interests include biomedical signal processing (interesting features and abnormalities found in bio-signals), theoretical modeling and simulation (breast cancer tissues), applied electronics (thermistors), process control and instrumentation, and embedded/VLSI systems. She has considerable research experience with one U.S. patent and two Indian patents for the development of thermal sensor-based breast cancer detection at its early stages together with the Centre for Materials for Electronics Technology (C-MET), an autonomous scientific society under Ministry of Electronics and Information Technology (MeitY), Government of India.

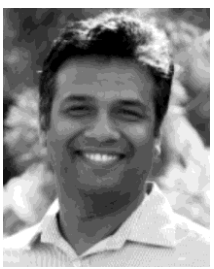

**Dr. SAMI AZAM** is currently a Leading Researcher and a Senior Lecturer at the Faculty of Science and Technology, Charles Darwin University, Casuarina, NT, Australia. He is also actively involved in Computer Vision, Signal Processing, Artificial Intelligence, and Biomedical Engineering research. He has several publications in peer-reviewed journals and international conference proceedings.

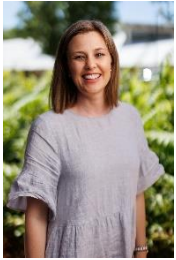

**Dr. GABRIELLE McCALLUM** is Senior Research Fellow, Senior Lecturer, and Program Leader of the Menzies' Child Health Respiratory team in Darwin. With a career spanning more than two decades in the Northern Territory, Dr. McCallum is dedicated to enhancing clinical outcomes for children susceptible to adverse lung health outcomes, primarily by addressing early and recurrent acute lower respiratory infections. Her approach involves conducting evidence-based research, developing culturally relevant educational materials, and translating research results into meaningful and culturally appropriate outcomes. Dr. McCallum's expertise has extended beyond national boundaries, with her contributions recognized nationally and internationally, including in regions such as New Zealand, Alaska, Malaysia, and Timor-Leste.

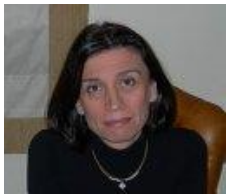

**Prof. EFTHYMIA ALEXOPOULOU** is a Professor of Pediatric Radiology at the National and Kapodistrian University of Athens Medical School. Her research focuses on the respiratory system, particularly in community-acquired pneumonia, complicated pneumonia, and magnetic resonance imaging in children and adolescents. Additionally, she has a keen interest in studying the impact of these conditions on the quality of life of her patients.

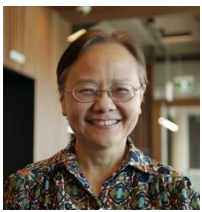

**Prof. ANNE CHANG** is a Senior Staff Specialist at the Queensland Children's Hospital, Brisbane. She leads the Cough and Airways Group at the Queensland University of Technology and is the Child Health Division's leader at Menzies in Darwin. She is a clinician recognized for her research contributions to evidence-based management and clinical care in pediatric cough, asthma, bronchiectasis, and Indigenous child lung health. Her original works include the world's first description of protracted bacterial bronchitis and international multicenter trials involving children with bronchiectasis. She has been a NHMRC practitioner fellow since 2004 and has published over 580 articles. Her primary interests are undertaking clinical research that improves Indigenous health management, cough, and suppurative lung disease in children.

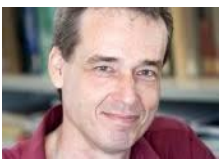

**Prof. FRISO De BOER** is a Professor of Engineering with the Faculty of Science and Technology, Charles Darwin University, Casuarina, NT, Australia. His research interests include signal processing, biomedical engineering, and mechatronics.